\documentclass[12pt,a4paper,oneside]{article}
\usepackage[top=1in,bottom=1in,left=.5in,right=.5in]{geometry}
\usepackage[margin=.5in,small]{caption}

\usepackage{authblk}
\usepackage{cite}
\usepackage{amsmath,gensymb,bigints}
\usepackage{epsfig}
\usepackage{graphics}
\usepackage{amsfonts, amssymb, upgreek}
\usepackage{textcomp}
\usepackage[english]{babel}
\usepackage{blindtext}
\usepackage{color}
\usepackage{braket}

\graphicspath{{img/}}

\providecommand{\abs}[1]{\vert #1\vert}

\newcommand{\bxi}{\boldsymbol{\xi}}

\newcommand{\tumble}{\mathtt{T}}
\newcommand{\run}{\mathtt{R}}

\newcommand{\const}{\mathop{}\!\mathrm{const}}
\newcommand{\upd}{\mathop{}\!\mathrm{d}}
\newcommand{\arsinh}{\mathop{}\!\mathrm{arsinh}}

\newcommand{\ex}{\mathrm{e}}

\newcommand{\Or}{\mathrm{O}}

\begin{document}

\title{A stochastic model for directional changes of
swimming bacteria}

\author[1]{G. Fier}
\author[1,2]{D. Hansmann\thanks{David.Hansmann@conicet.gov.ar}}
\author[1,2]{R. C. Buceta\thanks{rbuceta@mdp.edu.ar}}

\affil[1]{Instituto de Investigaciones F\'{\i}sicas de Mar del Plata, UNMdP and CONICET}
%\affil[2]{Departamento de Biolog\'{\i}a, FCEyN, Universidad Nacional de Mar del Plata}
\affil[2]{Departamento de F\'{\i}sica, FCEyN, Universidad Nacional de Mar del Plata}
\affil[{ }]{Funes 3350, B7602AYL Mar del Plata, Argentina}
%\date{Received: date / Revised version: date}

\maketitle

\abstract{In this work we introduce a stochastic model to describe directional changes in the movement of swimming bacteria. We use the probability density function (PDF) of turn angles, measured on tumbling wild-type {\sl E. coli}, to build a Langevin equation for the deflection of the bacterial body swimming in isotropic media. We have solved this equation analytically by means of the Green function method and shown that three parameters are sufficient to describe the movement: the characteristic time, the steady-state solution and the control parameter. We conclude that the tumble motion, which is manifested as abrupt turns, is primarily caused by the rotational boost generated by the flagellar motor and complementarily by the rotational diffusion introduced by noise. We show that in the tumble motion the deflection is a non-stationary stochastic process during times at which the tumbling occurs. By tuning the control parameter our model is able to explain small turns of the bacteria around their centres of mass along the run. We show that the deflection during the run is an Ornstein–Uhlenbeck process, which for typical run times is stationary. We conclude that, along the run, the rotational boosts do not exist and that only the rotational diffusion remains. Thus we have a single model to explain the turns of through a critical value that can explain the transition between the two turn behaviours. This model is the bacterium during the run or tumble movements, through a control parameter that can be tuned also able to explain in a very satisfactory way all available statistical experimental data, such as PDFs and average values of turning angles times, of both run and tumble motions.
}

%\pacs{02.50.-r, 05.10.Gg, 68.35.Ct, 89.75.Da}
%\submitto{\JPA}
%\maketitle

%\begin{keyword}
%bacterial swimming \sep run-tumble \sep e.coli \sep stochastic model  
%\end{keyword}

\section{Introduction}

Bacterial systems, among other microorganism systems, have the property of absorbing energy from their environment and storing it internally. The partial conversion of their internal energy into kinetic energy results in different specific movements of the individual active agents. This mechanism is absent in colloidal systems that are in thermal equilibrium, which are composed of passive particles that perform Brownian motion with an average speed tending to zero at very long times. Instead, self-propelled microorganisms (SPMs) without taxis (or directed motion) are active agents that are far from equilibrium at long times, changing incessantly between different metastable equilibrium states. In contrast, SPMs, under taxis in response to external stimuli (chemical, radiation, thermal, or magnetic, among others), move along the gradient of a guiding field with a nonzero drift speed \cite{Eisenbach2004}. The different kinds of bacterial movements have been classified by J\o{}ren Henrichsen \cite{Hen1972} and include individual movements ({\sl e.g.} swimming), collective movements ({\sl e.g.} flocking or swarming), as well as movements which occur individually and collectively ({\sl e.g.} gliding or twitching). The first studies concerning bacterial motions involving statistical mechanics were performed in the 1970s on swimming {\it E. coli} \cite{Berg1971,Berg1972,Berg1973} and build the base for the physics of microorganisms as a branch of soft condensed matter.

Flagellate SPMs (including species of bacteria, algae, protozoa, sperms, {\sl etc.}) have developed efficient mechanisms to move in bulk fluids or on moist surfaces between thin layers of fluid \cite{Pedley1992}. Commonly, the body of a motile flagellated bacterium (MFB) behaves like a rigid body, its basic movements are translations and rotations, while its deformations are negligible. Both, the translational and the rotational degrees of freedom can be reduced by the constraints which are imposed by the geometry of the medium and/or by the interactions between neighboring congeners. Often bacteria develop strategies (quorum-sensing, surfactant secretion, or other) to explore and colonize the resource rich environments (nutrients, temperature, oxygen, or other) and, thus, facilitate their development \cite{Purcell1977, Lauffenburger1991, Webre2003}. The different bacterial species show characteristic movement patterns ({\sl e.g.} run–tumble, run–reverse or run–reverse–flick) which can also depend on the density of the bacterial colony. MFB swim or swarm when they rotate their helical flagella ({\sl e.g. Escherichia coli} or {\sl Salmonella typhimurium}) each attached by a joint to a reversible rotary motor \cite{Berg1973}. A bacterium is pushed to the front when its flagellum or flagella (forming a bundle) turn with a definite chirality. The run of the bacterium is essentially determined by the anisotropy of the hydrodynamic friction generated by the slender body of the flagella, allowing a drag-based thrust \cite{Lauga2009}. 

\textit{E.\,coli} inoculated in a stimulus free environment where taxis effects can be neglected, shows a movement pattern composed of two alternate steps: run and tumble. During the run the bacterium moves forward with slight fluctuations in orientation and speed. During the tumble the bacterium stops and deflects its body with an abrupt turn in its orientation. Between the end of the tumble (run) and the start of the run (tumble) the bacterium undergoes a transition between two metastable equilibrium states, through internal mechanisms that are not yet fully understood \cite{Tu2005}. Viewed from behind, during the run, the flagella bundle of {\it E.\,coli} rotates counterclockwise (CCW) and during the slowdown (with a reverse thrust) the flagella rotate clockwise (CW) \cite{Macnab1977}. The change from CCW to CW rotation unbundle one or more filaments \cite{Turner2000}, causing a turn of the bacterium body around its center of mass \cite{Darnton2007}. It is commonly accepted that the center of mass of the bacterium does not move (or moves insignificantly \cite{Turner2000}) during the tumble and that only the direction of swimming changes. After tumbling, the bacterial motor switches from CW to CCW and all filaments form a new bundle which generates a forward thrust in the new direction. Thus, the swim motion of {\it E.\,coli} is reduced to two alternating stages called run-and-tumble, which have been studied as separate movements \cite{Condat2005,Saragosti2012} and in their entirety \cite{Zaburdaev2015}. A large number of experimental results are available for theoretical studies of run and tumble movements. This is particularly the case of {\it E.\,coli} experiments, in fact {\it E.\,coli} is one of the best studied microorganisms in regard to both its genomic and its biochemical processes \cite{Berg2004,Eisenbach2004}. 

Based on its simple dynamics, a swimmer bacterium can be characterized at a given time $t$ by its velocity $\mathbf{v}(t)$ and position $\mathbf{r}(t)$ within a three-dimensional (3D) reference frame. Taking into account that each tumbling motion might be performed in a different two-dimensional (2D) plane it is only possible to describe one sequence run-tumble-run in the same plane. The use of a reference frame with two coordinate axes on the tumble plane, in which the velocity \mbox{$\mathbf{v}=v(t)\,\mathbf{e}(t)$} has two components $(v_x,v_y)$ in Cartesian coordinates or $(v,\psi)$ in polar coordinates is the common 2D approach in order to treat the run and tumble movements separately \cite{Schienbein1993}. Assuming that the velocity $\mathbf{v}(t)$ is a two-dimensional continuous-time stochastic process, whose statistical properties depend on the studied system, the displacement of the center of mass of the bacterium during time $[t_0,t]$ is given by \mbox{$\mathbf{r}(t)-\mathbf{r}(t_0)=\int_{t_0}^t\mathbf{v}(t')\,\upd t'$}. The velocity correlation \mbox{$\langle\mathbf{v}(t')\cdot\mathbf{v}(t'')\rangle$} and the mean-squared displacement (MSD) \mbox{$\langle\abs{\mathbf{r}(t)-\mathbf{r}(t_0)}^2\rangle$} for run or tumble movements can be conveniently described in the reference frame too \cite{Taktikos2013}. 
A simple 2D projection of the position and the velocity of swimmer bacterium can be obtained by single-cell tracking based on the standard imaging microscopy. More detailed information can be obtained by 3D tracking, which requires special equipment and might be limited with respect to the statistical precision by the number of averages achieved \cite{Berg1971,Berg1972,Drescher2009}. Current devices use two identical orthogonal imaging assemblies that combine 2D projections to track 3D movements of the bacteria. A complementary 3D tracking method is the use of dynamic differential microscopy. But this method is used to characterize the motions of an entire population rather than movements of individual cells, by analysing temporal fluctuations of the particle density on different length scales {\sl via} image processing \cite{Wilson2011,Martinez2012}. The swimming speed distribution, fraction of motile cells, and diffusivity have been measured for {\sl E.\,coli} using these techniques. 

It is quite common to describe the dynamics of microorganisms in terms of stochastic differential equations (or Langevin equations). This is because of the rough similarity, originally observed, of the microscopic cell motion to Brownian motion \cite{Selmeczi2007}. Moreover, the observation of the tendency of cells to maintain their direction of movement during a characteristic time has led to the use of the idea of persistent random walks to describe the trajectories \cite{Romanczuk2012}. Langevin equations have been used to describe the stochastic motion of cells based on experimental observations \cite{Amselem2012, Selmeczi2005}. Based on this, the Langevin equations may include terms for self-propelling forces as well as external forces and random forces or torques, referred to as noise. Noise typically includes all fast variables of the system, {\sl e.g.} events that occur within very small time scales compared to the time-scale of the analysed process, like collisions between the microorganism and the surrounding medium, or intracellular processes involved in locomotion. 

The SPMs are usually characterized as active agents (or individuals moving actively gaining energy from the environment) or more precisely active Brownian particles (ABPs) \cite{Schimansky1995}. Fluctuations affecting the movement of each active agent may be due to internal or environmental processes. These systems can be effectively described as introducing a dissipation force $-f(\mathbf{r},\mathbf{v})\,\mathbf{e}(t)$ which points in the direction of movement. Simple models of ABP in homogeneous media have been studied in detail \cite{Schienbein1993, Steuernagel1994, Schimansky1995, Riethmuller1997, Klimontovich1994, Mikhailov2002, Schweitzer2003} using a velocity dependent dissipation force with intensity $f=f(\mathbf{v})$. Moreover, the existence of energy sources or nutrients can be modeled by a force $-\nabla U(\mathbf{r})$, where $U$ an attractive potential. The corresponding Langevin equation of such an ABP system is \mbox{$\dot{\mathbf{v}}=-f\,\mathbf{e}-\nabla U+\bxi(t)$}, where the noise components of $\bxi$ ({\it e.g.} $\xi_v$ and $\xi_\psi$) are Gaussian white noises, and it is easy to show that \mbox{$\upd E/\upd t=-f\,v+\bxi\cdot\mathbf{v}$}, where $E$ is the mechanical energy of the system. Assuming noise with correlation \mbox{$\langle\bxi(t)\cdot\bxi(t')\rangle=2\,D\,\delta(t-t')$} we obtain \mbox{$\langle\bxi\cdot\mathbf{v}\rangle=2 D$}, where $D$ is the noise intensity, and with small noise intensity we obtain \mbox{$\langle\upd E/\upd t\rangle\simeq-f\,v\,$}. All ABP models show that the system dissipate energy with $f>0$ for high speeds and show active friction (converting, partially, stored energy into kinetic) with $f<0$ for low speeds. In addition, the fluctuation-dissipation relation valid to Brownian particles becomes invalid to ABPs \cite{Joanny2003}.

Swimmer bacteria as well as other active agents, have a preferential orientation (or polarity), referred by the heading unit vector $\mathbf{e}(t)$, {\it e.g.} for {\it E.\,coli} the orientation from the tail to the head is chosen, which allows characterizing the persistence of movement. Not always the orientation coincides with the direction of movement. The speed $v(t)$ can be positive or negative according to the bacteria from moving forward or backward, respectively. Nonetheless, due to the impact of noise, swimming bacteria do not follow a straight line during their run movements. In the case of {\it E.\,coli} (wild type), that `runs' $\approx\!1\,\mathrm{sec}$, the noise introduces deflections (or orientation changes) with mean lateral turns of $\approx\!23\degree$ with respect to the mean direction \cite{Berg1972}. The extent of these straight-line deflexions depends on the runtime and as experiments with swimming bacteria shows the runtime (as well as the tumble time) is not constant but a random variable. Actually the runtime follows an exponential distribution \cite{Berg1972} or a power-law distribution \cite{Korobkova2004}. The tumble time, which also follows an exponential distribution, is typically an order of magnitude smaller than the runtime \cite{Berg1972}. However, some theoretical investigations use tumble-time distributions with the power-law behaviour \cite{Kafri2008}. In each tumble the bacterium undergoes reorientation with a distribution (of tumble angles) that is characteristic of each bacterial species and strain. To the best of the authors’ knowledge the first tumble-angle distribution (TAD) was measured by Berg and Brown (BB) for swimming {\it E.\,coli} \cite{Berg1972}. More recently studies deal with the TAD of pseudopod eukaryotic cells, \textit{e.g.} {\it Dictyostelium discoideum} \cite{Liang2008, VanHaastert2010}. 

In this paper, we address the stochastic dynamics of turn angles corresponding to run and tumble motions based on BB's TAD data \cite{Berg1972}. They determine, among other observables, the mean change in the bacterial direction from run to run, the mean change in the bacterial direction during runs, the mean tumble time, the mean runtime and, based on more than 1100 events, the tumble-angle distribution $P(\psi_\tumble)$. With the same aim as us, Saragosti {\sl et al.} \cite{Saragosti2012} proposed in 2012 a rotational diffusion process to model the tumble movement of {\it E.\,coli}. In contrast to Saragosti’s work, we assume that the tumble motion is an active stochastic process of the bacteria rather than a pure diffusion process. Taking into account the BB's {\it E. coli} TAD data as a starting point we propose in Section~\ref{sec:two} a Fokker-Planck equation for the stochastic process $x(t)=\cos\psi(t)$, with turn-angle $\psi$ and deflection $x$, both at time $t$. We show that the PDF of the deflection $x$ (for all $t$) is derivable from an equilibrium potential $U(x)$, where $-U'(x)$ is the drift term of the Langevin equation. We fit our theoretical PDF to the BB's TAD data using only three free parameters whose physical meaning we analyse.
We study the model for the tumble motion in Section~\ref{sec:three} and we obtain first the deterministic solution and then the stochastic solution of the Langevin equation. These solutions show that the deflection as a function of time is a stochastic process reduced to three parameters: the steady-state solution, the characteristic time, and the control parameter. We show that the deflection at tumble times is a nonstationary process and we find the mean deflection and the variance of the process. In Section~\ref{sec:four} we show that the proposed Langevin equation offers a solution for the deflections during the run motion, being a Ornstein–Uhlenbeck process that becomes stationary for characteristic runtimes. In Section~\ref{sec:five} we show how the deflections of the run and tumble motions are linked together based on experimental results, such that confirm a single model for both motions.
Finally, we discuss and interpret out main results in a biophysical context.

\section{Stochastic turn model\label{sec:two}}

The tumble motion is usually described in terms of two random variables, the tumble angle $\psi_\tumble$ and the tumble time $t_\tumble$. The tumble angle $\psi_\tumble$ is defined as the direction-change angle between the end of a run and the start of the following. In the present work, we use the tumble deflection \mbox{$x_\tumble=\cos\psi_\tumble$} (with $\abs{\psi_\tumble}\le \pi$) instead of the tumble angle for reasons of mathematical convenience. Previous studies have shown \cite{Saragosti2012} that the expansions in terms of Legendre polynomial of $x_\tumble$ fit the BB’s experimental data very well. The random variables $x_\tumble$ and $t_\tumble$ are completely characterized by the joint PDF \mbox{$P(x_\tumble,t_\tumble)$} which, based on experimental data, is not available. Usually the deflection PDF \mbox{$P(x_\tumble)=\int_0^{+\infty} P(x_\tumble,t_\tumble)\upd t_\tumble$\;} is determined measuring the experimental tumble-angle PDF $P(\psi_\tumble)$ (referred often to as TAD) without any further consideration of the tumble time $t_\tumble$. Complementary, the tumble-time PDF \mbox{$P(t_\tumble)= \int_{-1}^{1} P(x_\tumble,t_\tumble)\upd x_\tumble$\;} is measure without considering the deflection $x_\tumble\,$. Several authors have shown \cite{Berg1972,Korobkova2004} that tumble-time PDF follows an exponential behaviour, {\sl i.e.} $P(t_\tumble)\sim \ex^{-\zeta t_\tumble}$ (with $\zeta>0$). A well-established theoretical approach is the study of dynamics as a stochastic (or time-dependent random) process. In the present case, the stochastic process is the deflection by turning \mbox{$x(t)=\cos\psi(t)$} or, alternatively, in terms of heading unit vector as \mbox{$x(t)=\mathbf{e}(t_0)\!\cdot\!\mathbf{e}(t_0+t)$}, where $t_0$ is the initial time of the motion. The stochastic variable $x$, which refers to a single bacterium, is defined in the real interval $[-1,1]$. The joint PDF $P(x_\tumble,t_\tumble)$ of the random variables may be derived from the one-dimensional PDF of the stochastic process $x(t)$ named here $p(x,t)$ though \mbox{$P(x_\tumble,t_\tumble)=\int_{-1}^{1}\int_0^{+\infty} \delta(t-t_\tumble)\delta(x-x_\tumble)\,p(x,t)\upd t\upd x$\,}, where $\delta$ is the Dirac delta function. 
\begin{figure}
\begin{center}
\includegraphics[scale=0.5]{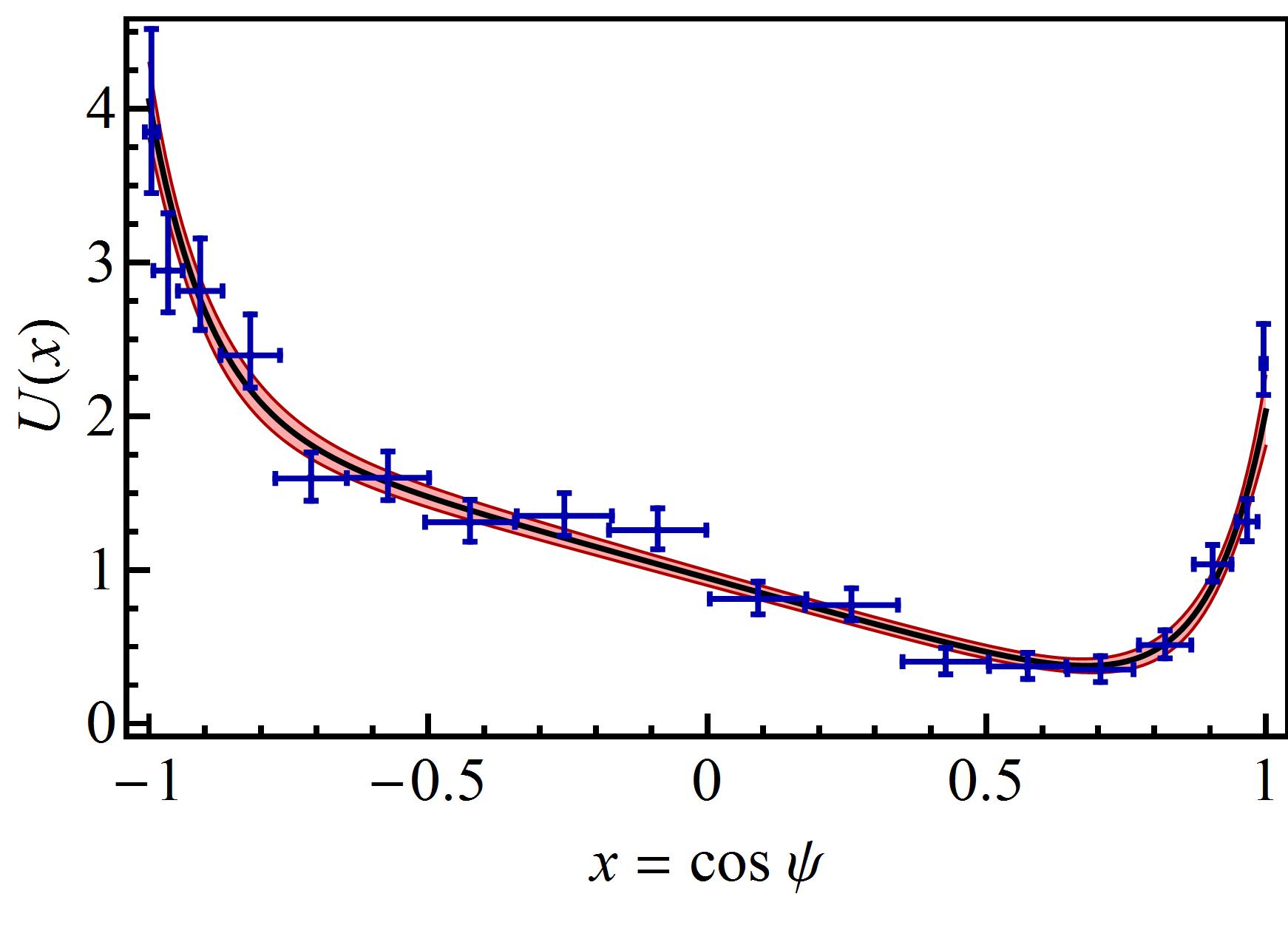} \hspace{.5cm}
\includegraphics[scale=0.5]{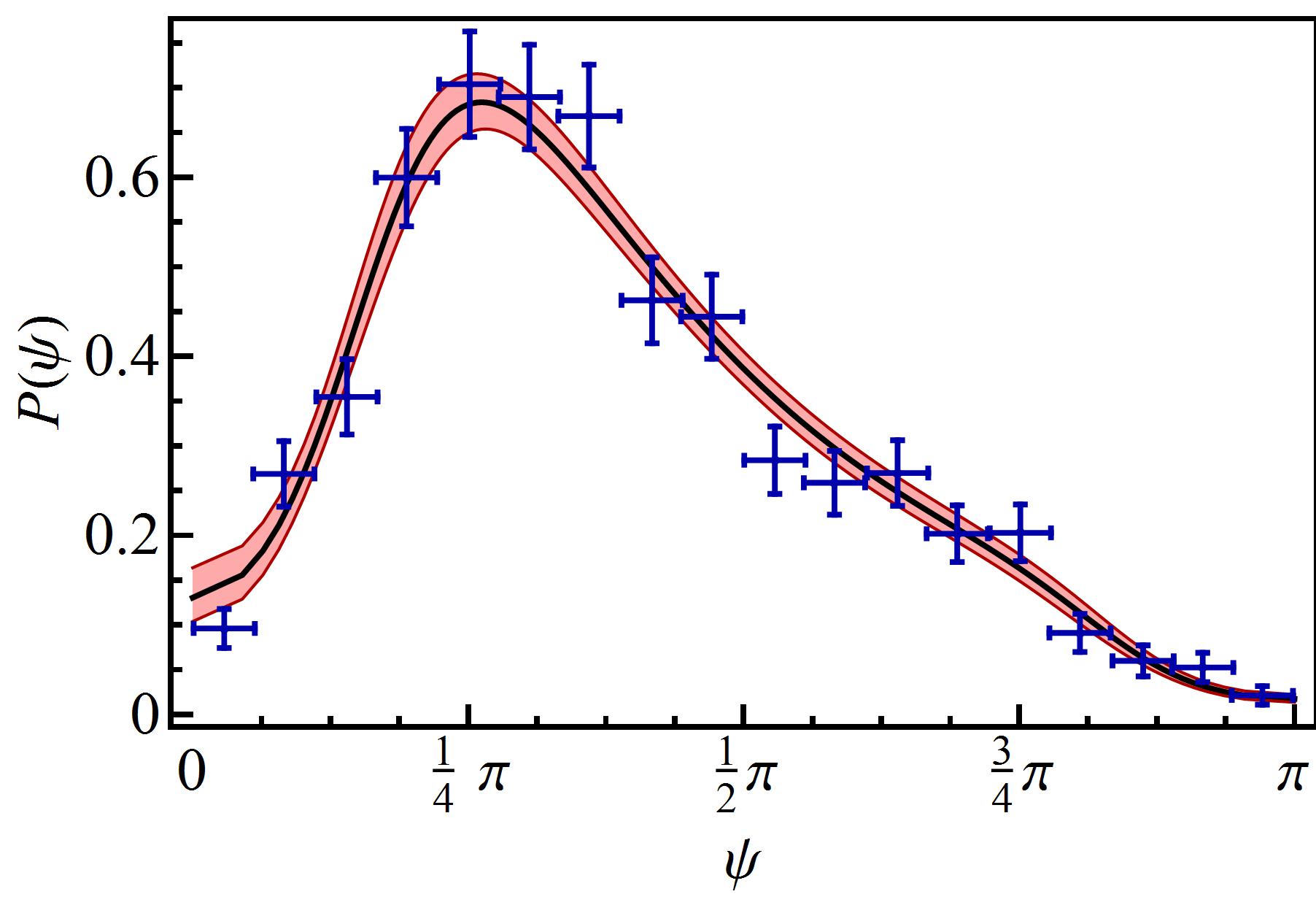}
\end{center}
\caption{(Color online) Left: The plot shows the potential $U(x)$ as function of the deflection $x=\cos\psi$ (black solid line), fitted to the TAD data of Berg and Brown \cite{Berg1972} (blue points with error bars). Around the fitted function the confidence interval of the fit is shown (pink shaded region). From the fit we obtain the parameters $\mu/D=1.00\pm 0.01$, $\nu/D=(7.41\pm 1.05)\times 10^{-4}$, and $\delta=9.74\pm 0.14$ (dimensionless). Right: The plot shows the tumble-angles PDF $P(\psi)$ (black solid line) modeled with equation~(\ref{tad}) using the fitted potential $U(x)$ shown to the left. The shaded region and error bars represent the confidence interval of the fit and the experimental uncertainties, respectively. Fit details: Based on the experimental TAD, we performed more than $10^5$ least-square adjustments to calculate the numerical values and uncertainties of the free parameters of our model $p(x)$, given by equation~(\ref{tad}), where for each adjustment the set of experimental data were randomly set assuming that (a) the measured angles are uniformly distributed within intervals of $10\degree$, and (b) the number of events fallen in each interval is Poisson-distributed.\label{fig:potential-Berg}}
\end{figure}

Assuming that $x(t)$ is a continuous Markov process we propose the following Fokker-Planck equation  
\begin{equation}
\frac{\partial\hspace{1ex}}{\partial t}p(x,t)= -\frac{\partial\hspace{1ex}}{\partial x} \bigg[ K_1(x) -\frac{\partial\hspace{1ex}}{\partial x}K_2(x)\biggr]p(x,t)\,\label{Fokker-Planck}
\end{equation}
with initial condition $p(x,0)=\delta(x-x_0)$, where $K_1$ and $K_2$ are the time-independent drift and diffusion coefficients, respectively \cite{Risken1989}. Assuming that \mbox{$p(x,t)\sim\ex^{-\kappa t}$} with $\kappa > 0$, the PDF of the deflection $x(t)$ becomes zero when the time goes towards infinite, {\sl i.e.} \mbox{$\lim_{t\to+\infty}p(x,t)=0$}. In consequence, for all $x\neq x_0\,$,
\begin{equation}
\biggl[K_1(x)-\frac{\partial\hspace{1ex}}{\partial x}\,K_2(x)\biggr] p(x) =\int_0^{+\infty}\!J(x,t)\,\upd t=\const\;,\label{int-current}
\end{equation}
where \mbox{\;$J(x,t)=(K_1-\frac{\partial\hspace{1ex}}{\partial x}\,K_2)\,p(x,t)$\;} is the probability density current of equation (\ref{Fokker-Planck}) and 
\begin{equation}
p(x)=\int_0^{+\infty}p(x,t)\,\upd t\nonumber
\end{equation}
is the PDF of the deflection $x=x(t)$. The PDF of deflection $x_\tumble$ can be recovered by means of $P(x_\tumble)=\int_{-1}^1 p(x)\delta(x-x_\tumble)\upd x$. We assume the constant of equation~(\ref{int-current}) is equal to zero in order to obtain the {\sl ansatz}
\begin{equation}
p(x)=\mathcal{N}\,\ex^{-U(x)}\;,\label{tad}
\end{equation}
where $\mathcal{N}$ is the normalization and the potential is
\begin{equation}
U(x)=\ln\,[K_2(x)]-\int^x\frac{K_1(x')}{K_2(x')}\,\upd x'\;.\label{eq:potential}
\end{equation}
Assuming a stochastic process with additive noise, we propose a constant diffusion coefficient $K_2(x)=D$ and a drift coefficient $K_1(x)=-D\,U'(x)$. Based on equations~(\ref{tad}) and (\ref{eq:potential}), the proposed potential is
\begin{equation}
U(x)=U_0-\frac{1}{D}\Bigl[\mu\,x-\nu\,\cosh(\delta x)\Bigr]\;,\label{ec1}
\end{equation}
where all constants are positive real numbers. The {\sl ansatz} of equation~(\ref{tad}) is validated empirically by the good agreement between the fitted model and the experimental data. The potential fits very well with the experimental \textit{E. coli} data of Berg and Brown \cite{Berg1972} as shown in Figure~\ref{fig:potential-Berg}. Notice that the potential of equation~(\ref{ec1}) is defined with only 3 parameters: $\mu/D$, $\nu/D$ and $\delta$, while $U_0$ is eliminated by normalization. Without loss of generality we set $\mu=D$ since the fit of the coefficients $\mu$ and $D$ yield $\mu/D\approxeq 1$ (see Figure~\ref{fig:potential-Berg}). Consequently the corresponding  Langevin equation of the tumble deflection $x(t)$ is
\begin{equation}
\dot{x}=K_1(x)+\eta(t)\;,\label{Langevin-eq}
\end{equation}
where the drift coefficient is
\begin{equation}
K_1(x)=\mu-\nu\delta\,\sinh(\delta x)\;,
\end{equation}
and $\eta=\eta(t)$ is additive Gaussian white noise with the mean $\bigl\langle\eta(t)\bigr\rangle=0$ and the correlation
\begin{equation}
\bigl\langle\eta(t)\,\eta(t')\bigr\rangle=2D\,\delta(t-t')\,,\label{noise-correlation}
\end{equation} 
where $D$ is the noise intensity.

\section{Tumble motion\label{sec:three}}

\begin{figure}[t]
\begin{center}
\includegraphics[scale=0.5]{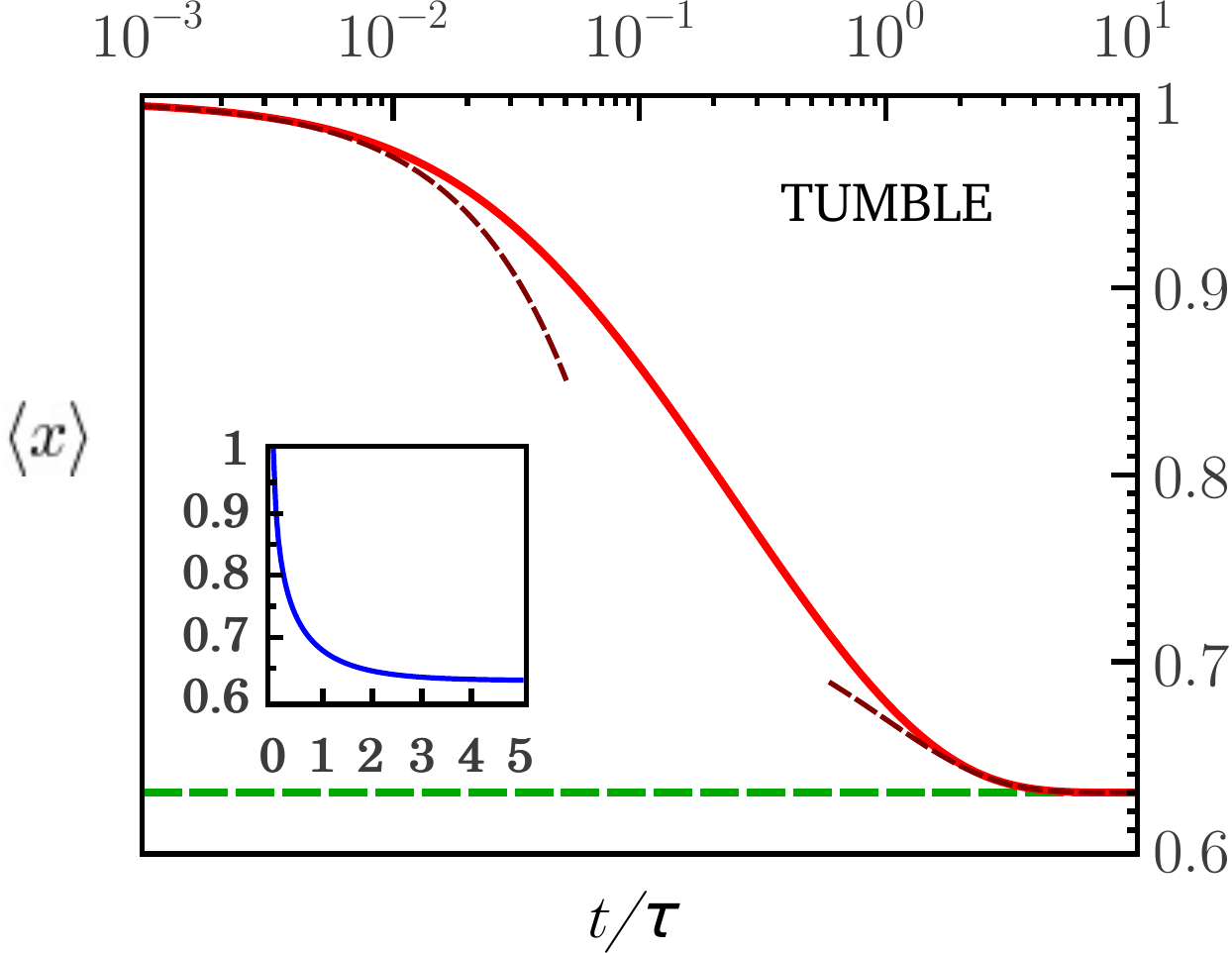}
\end{center}
\caption{(Color online) Log-linear plot of the mean deflection $\langle x\rangle$ {\it vs} the adimensional time $t/\tau$, for the tumble motion obtained from equation~(\ref{mean-deflection}) (red solid line). Inside {\sl idem} the linear plot (blue). For the tumble motion the curves are in good agreement with the deterministic solution given by equation~(\ref{x-vs-t}). Initially the deflection is $x_0=1$ and after a short transient $x$ goes asymptotically to $x_\mathrm{s}$ given by equation~(\ref{eq5}). The dotted curves illustrate equation~(\ref{x-vs-t-early}), valid for in the early regime $(t\ll\tau)$, and equation~(\ref{x-vs-t-equil}), valid in the late regime $(t\gg\tau)$. The data used in this plot are $\gamma=0.00663$ and $\delta=9.062$. The steady-state value is $x_\mathrm{s}\approxeq 0.630$ (calculated).  
\label{fig:0}}
\end{figure}

\paragraph{Deterministic solution.} In order to determine the meaning of phenomenological constants we integrate the system assuming that the noise $\eta(t)=0$. Integrating the deterministic equation
\begin{equation}
\dot{X}=\mu-\nu\delta\,\sinh(\delta X).\;\label{deterministic-eq}
\end{equation}
on the interval $[0,t)$, with initial condition $X(0)=1$, yields
\begin{equation}
\Lambda(X)=\frac{\gamma\,\ex^{-\delta X}+1-\sqrt{1+\gamma^2}}
{\gamma\,\ex^{-\delta X}+1+\sqrt{1+\gamma^2}}=\Lambda_1\,\ex^{-t/\tau}\;,\label{eq3}
\end{equation}
where $\Lambda_1=\Lambda(1)\,$,\; $\gamma=\nu\delta/\!\mu\,$,\; and 
\begin{equation}
\tau=\frac{1}{\mu\,\delta\sqrt{1+\gamma^2}}\label{tau}
\end{equation}  
is the characteristic time of the turn. Taking the \textit{E. coli} tumble data into account one can estimate that $\gamma\ll 1$ (see Figure~\ref{fig:potential-Berg}).  
In consequence, the zero order expansion in $\gamma$ of the characteristic time is $\tau\simeq(\mu\delta)^{-1}$. In addition we infer from the fit of our model to the experimental data of BB that $\mu=D$. Without loss of
generality, we chose $D=1$ and calculated $\tau\approx 0.10$, which is close to experimental tumble times of \textit{E. coli}. Setting $K_1(x_\mathrm{s})=0$ (or equivalently $U'(x_\mathrm{s})=0$) we obtain the steady-state solution 
\begin{equation}
x_\mathrm{s}=\frac{1}{\delta}\arsinh\Bigl(\frac{1}{\gamma}\Bigr)<1\;.
\label{eq5}
\end{equation}
The deterministic solution depends on three parameters $\{\gamma,\delta,\mu\}$  from which the two physical quantities $\{\tau,x_\mathrm{s}\}$ are derived. The solution of the equation~(\ref{deterministic-eq}), with the initial condition $X(0)=1$, is
\begin{equation}
X(t)=-\frac{1}{\delta}\,\ln\biggl[\frac{\sqrt{1+\gamma^2}}{\gamma}\biggl(\frac{1+\Lambda_1\,\ex^{-t/\tau}}{1-\Lambda_1\,\ex^{-t/\tau}}\biggr)-\frac{1}{\gamma}\biggr]\,,\label{x-vs-t}
\end{equation} 
which is discussed in Figure~(\ref{fig:0}).
For short times ($t\ll\tau$), we observed an exponential change of the bacterial orientation, starting from the unstable equilibrium ($X=1$ or $\psi=0$). Using the auxiliary function $y(t)=\ex^{-t/\tau}$, we expand the Taylor series of $X(t)$ around $y=1$ (or equivalently around $t=0$) up to first order and obtain the deflection close to the unstable equilibrium ($X\lessapprox 1$) given by
\begin{equation}
X(t)\simeq 1+\alpha_1\bigl(\ex^{-t/\tau}-1\bigr)\;,\label{x-vs-t-early} 
\end{equation}
where $\alpha_1\simeq -4\,\Lambda_1/[(4\,\Lambda_1+\gamma^2)\delta]$ assuming that $-4\Lambda_1\lessapprox\gamma^2\ll 1$. After some time, the bacterial orientation approaches a stable equilibrium at deflection $X=x_\mathrm{s}$ for $(t\gtrsim\tau)$. Expanding the Taylor series of $X(t)$ around $y=0$  (or equivalently for $t\to +\infty$) up to first order, we find an exponential approach to a stable equilibrium ($X\gtrapprox x_\mathrm{s}$) given by
\begin{equation}
X(t)\simeq x_\mathrm{s}+\alpha_0\,\ex^{-t/\tau}\;,\label{x-vs-t-equil}
\end{equation}
where $\alpha_0\simeq -4\,\Lambda_1/(\gamma^2\delta)$ assuming that $\gamma\ll 1$. The two analytical approximations (for short and long times) are connected by the exact deterministic solution as it is shown in Figure~(\ref{fig:0}). Contrary to expectations, this transient regime connotes a slowdown of the bacterial turn (clearly recognizable in the outer plot of Figure~(\ref{fig:0})). 

\paragraph*{Stochastic solution.}The solution of Langevin equation~(\ref{Langevin-eq}) is 
\begin{equation}
x(t)=x_\mathrm{s}+(x_0-x_\mathrm{s})\,G(t,t_0)+\int_{t_0}^t\eta(s)\,G(t,s)\upd s\;,\label{sol-Green}
\end{equation}
where 
\begin{equation}
G(t,t')=\frac{X(t)-x_\mathrm{s}}{X(t')-x_\mathrm{s}}\,H(t-t')\;,\label{Green-function-ex}
\end{equation}
is the Green's function of the problem, $X(t)$ is the deterministic solution given by equation~(\ref{x-vs-t}), and $H$ is the Heaviside step function defined as $H=1$ if $t\ge t'$ and $H=0$ otherwise. Taking into account (from the fit-parameters values shown in Figure~\ref{fig:potential-Berg}) that $-4\Lambda_1/\gamma^2\lessapprox 1$, we expand equation~(\ref{Green-function-ex}) around $\gamma=0$ to obtain
\begin{equation}
G(t,t')=\frac{\ln(1-\beta\,\ex^{-t/\tau})}{\ln(1-\beta\,\ex^{-t'\!/\tau})}\,H(t-t')+\Or(\gamma^2)\;,\label{Green-function-ap}
\end{equation}
where
\begin{equation}
\beta=-\frac{4\,\Lambda_1}{\gamma^2}\label{beta}
\end{equation}
is the control parameter of the turn motion. Assuming that the noise $\eta(s)$ of equation~(\ref{sol-Green}) is white noise with zero mean, the expectation value of $x(t)$ is
\begin{equation}
\langle x(t)\rangle =x_\mathrm{s}+(x_0-x_\mathrm{s})\,G(t,t_0)\;.\label{mean-deflection}
\end{equation}
The approximation of equation~(\ref{mean-deflection}) to zero-order in $\gamma$ agrees very well with the exact deterministic solution $X(t)$ given by equation~(\ref{x-vs-t}). Using equation~(\ref{sol-Green}) with correlated noise of equation~(\ref{noise-correlation}), the covariance \mbox{$r(t,t')=C[x(t),x(t')]=\bigl\langle[x(t)-\langle x(t)\rangle]\,[x(t')-\langle x(t')\rangle]\bigr\rangle$} is 
\begin{equation}
r(t,t')=2 D\int^{\min(t,t')}_{t_0}G(t,s)\,G(t',s)\,\upd s\;.\label{autocov}
\end{equation}
Using equation~(\ref{Green-function-ap}) one can show that \mbox{$G(t,t')\simeq\ex^{-\abs{t-t'}/\tau}\,$} if $\,\min(t,t')\gg\tau\,$, concluding  that at very long times the solution $x(t)$ of equation~(\ref{sol-Green}) describes an Ornstein-Uhlenbeck process. Substituting $u = \beta\,\ex^{-s/\tau}$ in the integrand (eq.~(\ref{Green-function-ap})) of equation~(\ref{autocov}) we explicitly obtain 
\begin{equation}
r(t,t')=2 D\tau\,\ln(1-\beta\,\ex^{-t/\tau})\,\ln(1-\beta\,\ex^{-t'\!/\tau})\bigintssss_{\beta\,\ex^{-[\min(t,t')/\tau]}}^{\beta\, \ex^{-(t_0/\tau)}}\;\frac{\upd u}{u\,\ln^2(1-u)}\;.
\end{equation}
In order to calculate the integral we approximate integrand expanding it in a Laurent series around $u = 0$. Taking in to account that the indefinite integral is 
\begin{equation}
\mathcal{I}(u)=\int \frac{\upd u}{u\ln^2(1-u)}=-\frac{1}{2u^2}+\frac{1}{u}+\frac{\ln u}{12}-\frac{u^2}{480}-\frac{u^3}{720}+\Or(u^4)\;,\label{integral}
\end{equation}
that $\ln(1-\beta\,\ex^{-t/\tau})\simeq\beta\,\ex^{-t/\tau}$ if $t\gg\tau$, and that $\min(t,t')=\frac{1}{2}(t+t')-\frac{1}{2}\abs{t-t'}$, one can show that
\begin{equation}
\lim_{t\to +\infty} r(t,t+T)=R(T)=D\tau\,\ex^{-\abs{T}/\tau}\;.\label{asym-autocorr}
\end{equation}
This limit, together with $\lim_{t\to +\infty}\langle x(t)\rangle=x_\mathrm{s}\,$, shows that for very long times the bacterium-turn process becomes stationary. 
The process, however, is not stationary for short times including typical tumble times $t_\tumble$ that are in the order of the characteristic time $\tau$. With $t_0=0$, the variance $v(t)=r(t,t)$ is
\begin{eqnarray}
v(t)\!\!&=&\!\!D\,\tau\biggl[1-\beta\,\ex^{-t/\tau}+ \biggl(\frac{t}{6\,\tau}+2\,\mathcal{J}(\beta)-\frac{13}{12}\biggr)\,\beta^2\,\ex^{-2\,t/\tau} +\biggl(\frac{t}{6\,\tau}+2\,\mathcal{J}(\beta)-1\biggr)\,\beta^3\,\ex^{-3\,t/\tau}\nonumber\\
&&\hspace{0.5cm}+\;\frac{11}{12}\biggl(\frac{t}{6\,\tau}+2\,\mathcal{J}(\beta)-\frac{708}{720}\biggr)\,\beta^4\,\ex^{-4\,t/\tau}
+\Or\bigl(\ex^{-5\,t/\tau} \bigr)\biggr]\;,\label{time-variance-tumble}
\end{eqnarray}
where $\mathcal{J}(\beta)=\mathcal{I}(\beta)-\frac{1}{12}\ln\beta$. The variance for the stationary limit $t\to+\infty$ is $v_\infty=D\tau=\delta^{-1}$. Equation~(\ref{time-variance-tumble}) confirms that the process is non-stationary for $t=t_\tumble\,$ as mentioned previously.
\begin{figure}
\begin{center}
\includegraphics[scale=0.5]{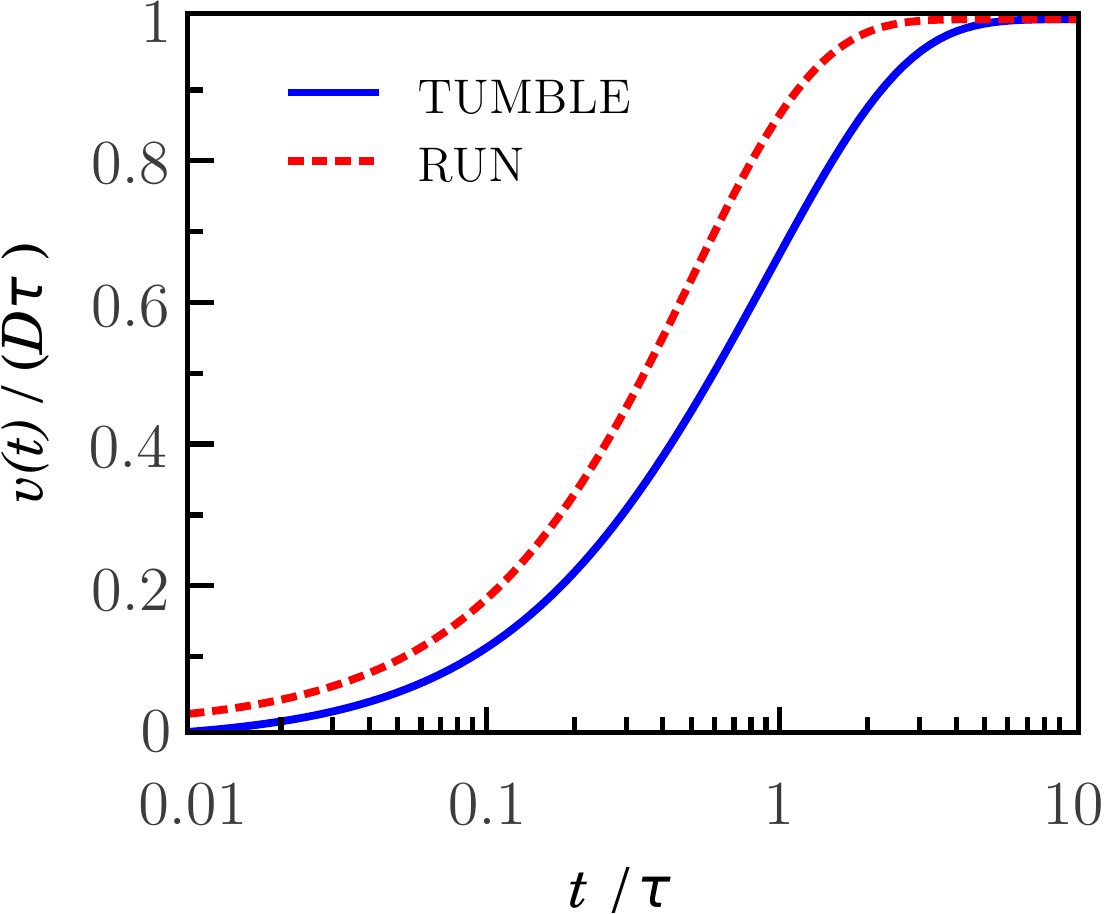} \hspace{2cm}
\includegraphics[scale=0.5]{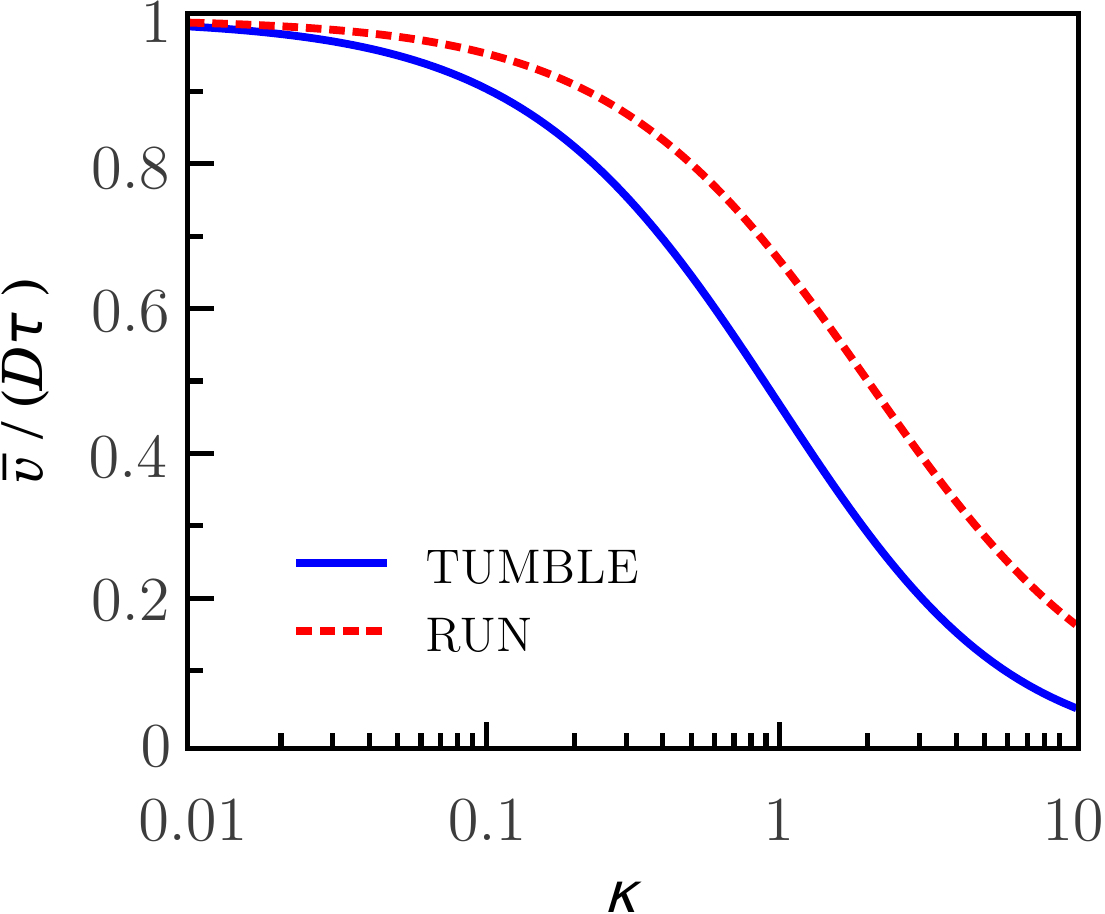}
\end{center}
\caption{(Color online) Both plots: Show tumble-motion data (blue solid line) and run-motion data (red dashed line). Left: The plot shows the time dependent behaviour of the variance $v(t)$ given by equations~(\ref{time-variance-tumble}) and (\ref{time-variance-run}) (dimensionless with the asymptotic value $D\tau$). Initially at $t_0 = 0$ the system is completely uncorrelated. On the contrary, for $t\gg\tau$ the system becoming stationary and correlates completely. For intermediate times $t= t_\tumble\approx\tau$, an intermediate response is observed which leads to the conclusion that the deflection $x(t)$ is a non-stationary process. Right: The plot shows the average variance $\bar{v}$ given by equations~(\ref{variance-average-tumble}) and (\ref{variance-average-run}) (dimensionless with $D\tau$) as function of $\kappa$, the exponent of the TTD. In both plots, the tumble-motion data (solid line) are plotted taking $\beta=0.965$.
\label{fig:variance}}
\end{figure}
This result seems natural taking into account that the turn of the bacterium ends at tumble times long before reaching the steady state, as we shows at left plot of Figure~\ref{fig:variance} (with solid line). Thus, the experimental TAD is measured over an ensemble of tumble times. The average variance of the tumble times is
\begin{equation}
\tilde{v}=\int_0^{+\infty}v(t_\tumble)\,P(t_\tumble)\upd t_\tumble\;,
\end{equation}
where $P(t_\tumble)$ is tumble time distribution (TTD). In accordance with the experimental results, we consider an exponential distribution $P(t_\tumble)=\mathcal{N}\,\ex^{-\kappa\,t_\tumble/\tau}$ and hence obtain the average variance 
\begin{eqnarray}
\frac{\tilde{v}}{D\,\tau}\!\!&=&\!\!1-\frac{\beta\,\kappa}{1+\kappa}+ \frac{\beta^2\kappa}{2+\kappa}\biggl(-\frac{1}{6\,(2+\kappa)}+2\,\mathcal{J}(\beta)-\frac{13}{12}\biggr)+\frac{\beta^3\kappa}{3+\kappa}\biggl(-\frac{1}{6\,(3+\kappa)}+2\,\mathcal{J}(\beta)-1\biggr) \nonumber\\ &&+\;\frac{\beta^4\kappa}{4+\kappa}\biggl(-\frac{1}{6\,(4+\kappa)}+2\,\mathcal{J}(\beta)-\frac{708}{720}\biggr)+\Or(\beta^5)\;.\label{variance-average-tumble}
\end{eqnarray}
The plot on the right side of Figure~\ref{fig:variance} (solid line) shows how far away is the average variance of the asymptotic value $D\tau$, as a function of parameter $\kappa$. The tumble time PDF  \mbox{$P(t_\tumble)=\sum_{j=1}^{+\infty}A_j\,\ex^{-\kappa_j\,t_\tumble/\tau}$} leads to a more accurate approach, which includes our approximation as a special case (where $A_j=\mathcal{N}\,\delta_{1j}$).

\section{Run motion\label{sec:four}}

Experimental observations of the swimming bacteria show that their run motion is not completely rectilinear, but that the bacteria perform random turns with angles $\psi(t)$ around their centres of mass that change their swimming directions. The random deflection $x(t)=\cos\psi(t)$ from a straight line during the run, is not always small and depends on the bacterial strain, body length, or other biological factors. In addition, the average of the deflection-angle absolute value $\langle\abs{\psi}\rangle$ and its uncertainties $\sigma_{\abs{\psi}}$ are relatively big too as reported for {\sl E. coli} ({\sl e.g.} $\abs{\psi}\approx 23 \degree\pm 23\degree$ for wild type)\cite{Berg1972}. Nevertheless experimental results suggest that most probable deflection angle at the end of a run is $\psi_\mathrm{s}\simeq 0\degree$. 
A steady state solution at $x=x_\mathrm{s}=1$ based on equations~(\ref{tad}) and (\ref{ec1}) requires that the potential $U(x)$, with $\abs{x}\le 1$,  has a global minimum at $x=1$. Equivalently, $x=1$ is a steady solution if
\begin{equation}
x^*=\frac{1}{\delta}\arsinh\biggl(\frac{1}{\gamma}\biggr)\ge 1\label{eq:xstar}
\end{equation}
exists, such that $U'(x^*)=0$. This condition might be satisfied with a proper choice of parameters $(\delta,\gamma)$. Notice that the parameter $x^*$ is an amount without physical meaning. In turn, the parameters $\delta$ and $\gamma$ are linked through the control parameter $\beta$. Inverting the equation~(\ref{beta}) we obtain
\begin{equation}
\delta(\beta,\gamma)=-\ln\biggl[\frac{\sqrt{1+\gamma^2}}{\gamma}\biggl(\frac{1-\frac{1}{4}\,\beta\,\gamma^2}{1+\frac{1}{4}\,\beta\,\gamma^2}\biggr)-\frac{1}{\gamma}\biggr]\;.\label{eq:delta}
\end{equation} 
\begin{figure}[t]
\begin{center}
\includegraphics[scale=0.53]{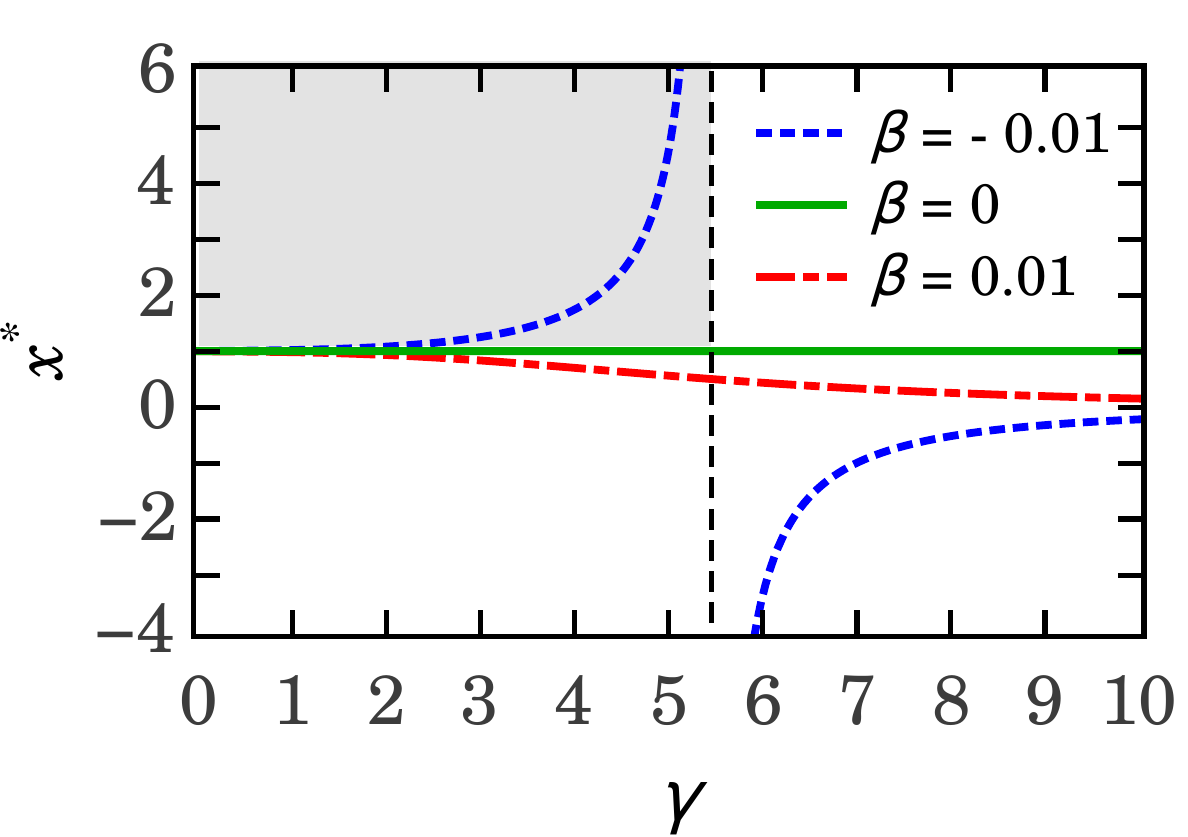}\hspace{1cm} 
\includegraphics[scale=0.5]{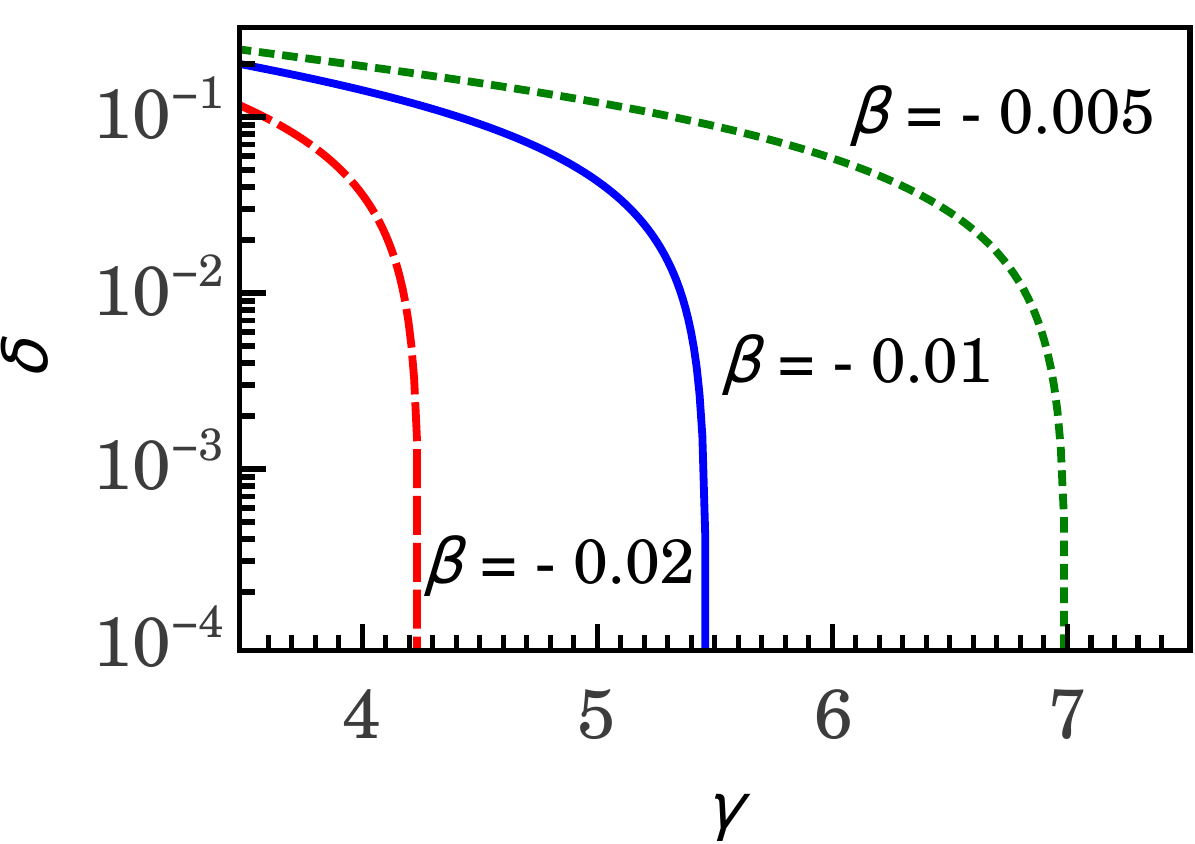} 
\end{center}
\caption{(Color online) Left: Plot of the parameter $x^*$ {\sl vs} the parameter $\gamma$ for several values of the control parameter $\beta$ close to zero. Notice that $x^*\ge 1$ when $\beta\lessapprox 0$ and $\gamma<\gamma_\mathrm{c}$ (where $x^*\to+\infty$ when $\gamma\to\gamma_\mathrm{c}^-$). All curves outside of the upper left quadrant (gray background) are not physical. Right: Plot of $\delta$ {\sl vs} $\gamma$,  from equation~(\ref{eq:delta}), for several values of the control parameter $\beta\lessapprox 0$. 
\label{fig:run-motion-1}}
\end{figure}
The plot shown on the left side of Figure~\ref{fig:run-motion-1} shows that the condition $x^*\ge 1$ holds when the parameters $\beta\lessapprox 0\,$ and \mbox{$\gamma<\gamma_\mathrm{c}$\,}, where $\gamma=\gamma_\mathrm{c}$ is the vertical asymptote of the $x^*(\gamma)$. Complementary, the plot shown on the right side of  Figure~\ref{fig:run-motion-1} shows the parameter $\delta$ as a function of the parameter $\gamma$, for several values of $\beta\lessapprox 0$.
\begin{figure}[t]
\begin{center}
\includegraphics[scale=0.5]{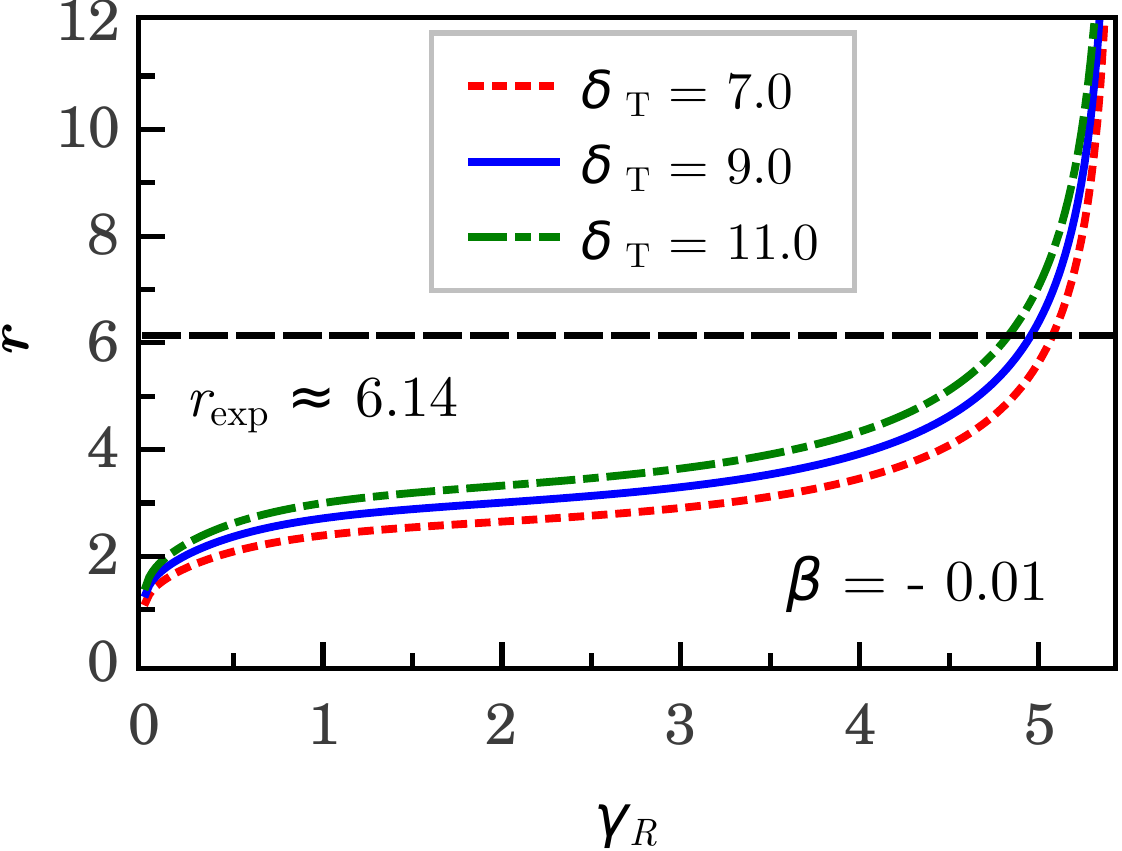}\hspace{1cm} 
\includegraphics[scale=0.44]{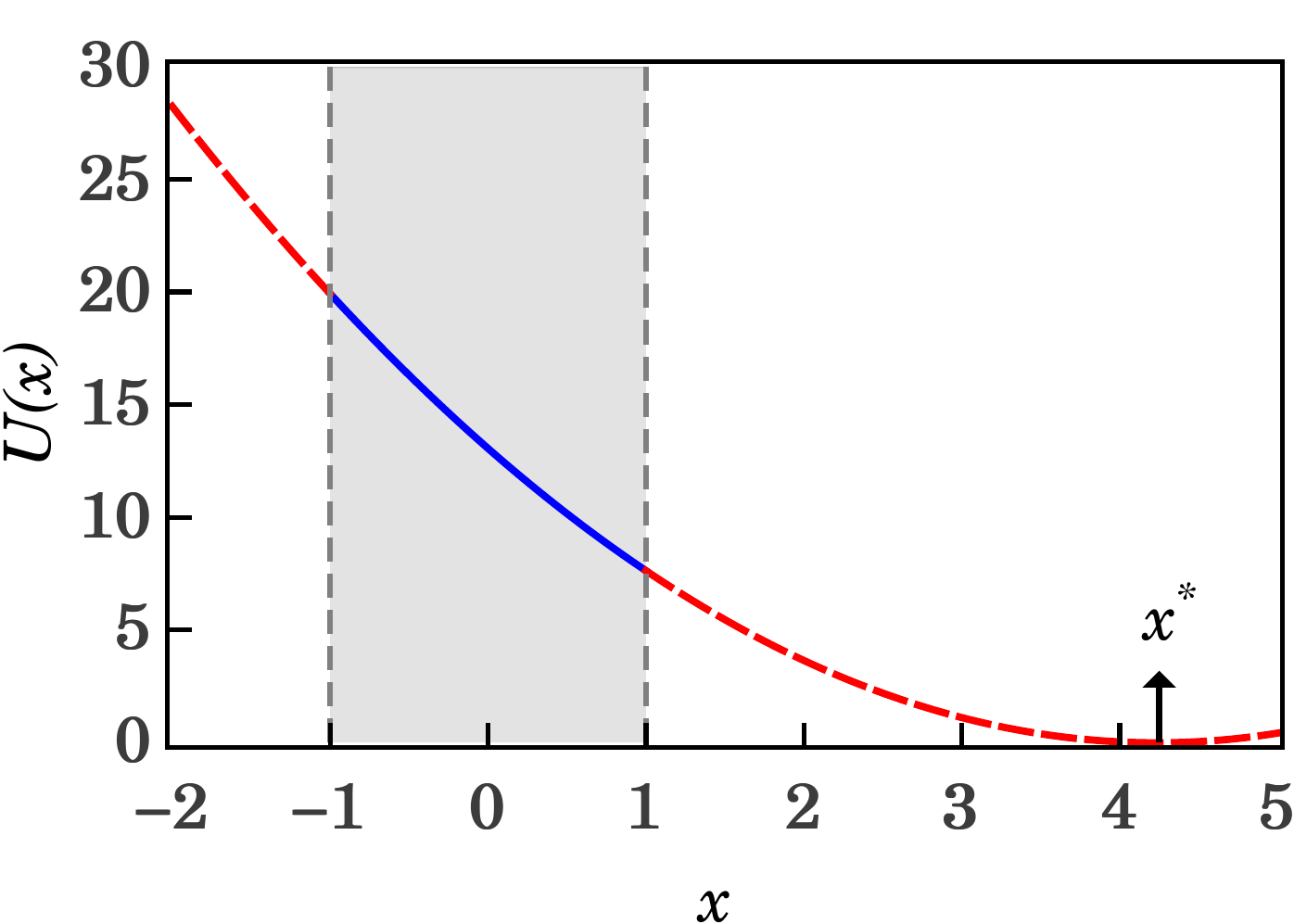}
\end{center}
\vspace{-2ex}\caption{(Color online) Left: Plot of the ratio $r$ (between the mean runtime and the mean tumble time) {\sl vs.} the parameter $\gamma_\run$ for different values of $\delta_\tumble$. One observes that at the intersection of the dashed line with the curves, where $r=r_\mathrm{exp}$, the values of $\gamma_\run$ are restricted ({\sl e.g.} for $r_\mathrm{exp}\approx 6.14$ and $\beta=-0.01$ we obtain $4.83\lesssim\gamma_\run\lesssim 5.08$ when $11\ge\delta_\tumble\ge 7$). Right: Plot of the potential $U(x)$ as function of the turn deflection $x$ of the run motion, defined in $\abs{x}\le 1$ (solid line, gray background). Outside of this interval the potential function $U(x)$ in non physical (red dashed lines). Within the interval $x>1$ we find $U'(x^*)=0$. The parameters used for this plot are $\beta=-0.01$, $\gamma=4.98$ and $r=6.15\,$, where $x^*\simeq 4.250$ and $\delta\simeq 0.0471$ are calculated using equations~(\ref{eq:xstar}) and (\ref{eq:delta}), respectively.
\label{fig:run-motion-2}}
\end{figure}
%Assuming for the run motion that the control parameter $\beta\approx 0$, it is easy to shows from equation~(\ref{beta}) that
%\begin{equation}\delta(\beta,\gamma)=\ln\biggl(\frac{\gamma}{\sqrt{1+\gamma^2}-1}\biggr)+\biggl[\frac{\gamma^2\,\sqrt{1+\gamma^2}}{2\,(\sqrt{1+\gamma^2}-1)}\biggr]\,\beta+\Or(\beta^2)\;.\label{eq:delta}\end{equation}

\paragraph{Stochastic solution.}
Setting $x_0=x_\mathrm{s}$ in equation~(\ref{sol-Green}), the deflection during the run motion is
\begin{equation}
x(t)=x_\mathrm{s}+\int_{t_0}^t\eta(s)\,G(t,s)\upd s\;,\label{sol-Green-run}
\end{equation}
which reflects that the deflections around the equilibrium are fully random. Taking into account that $\beta\lessapprox 0$ we can expand equation~(\ref{Green-function-ex}) around $\beta=0$. The corresponding Green's function is
\begin{equation}
G(t,t')=\ex^{-\abs{t-t'}/\tau}+\Or(\beta)\;.\label{Green-function-ap-run}
\end{equation}
This shows that the bacterial deflection from a straight line during the run motion is a stochastic process with solution given by equation~(\ref{sol-Green-run}) and with the Green's function given by equation~(\ref{Green-function-ap-run}) which describes an Ornstein-Uhlenbeck process. The covariance from equation~(\ref{autocov}) is
\begin{equation}
r(t,t')=D\,\tau\,\Bigl[\ex^{-\abs{t-t'}/\tau}-\ex^{-(t+t'-2t_0)/\tau}\Bigr]\;.
\end{equation} 
At very long times the system becomes a stationary Ornstein-Uhlenbeck process verifying the conclusion drawn from equation~(\ref{asym-autocorr}). This limit is satisfied for $\min(t,t')\gg\tau$. With $t_0=0$ the variance is 
\begin{equation}
v(t)=D\tau(1-\ex^{-2t/\tau})\label{time-variance-run}
\end{equation}
and, taking an exponential TTD $f(t)=\mathcal{N}\ex^{-\kappa\,t/\tau}$, the average variance is
\begin{equation}
\tilde{v}=D\tau\,\Bigl(\frac{2}{2+\kappa}\Bigr)\;.\label{variance-average-run}
\end{equation}
The plot shown on the left side of Figure~\ref{fig:variance} shows the variance of the deflection of the run motion (dashed line) as a function of time. Notice that the process becomes stationary for times longer than the characteristic time. The plot shows joint results of the run and tumble movements illustrating that the processes are stationary and nonstationary, respectively, at times which are longer than their characteristic times. The plot shown on the right side of Figure~\ref{fig:variance} shows that the average variance of the deflection of both run and tumble motions, for small values of the parameter $\kappa$, approach their asymptotic values $D\tau$.

\section{A single model for run and tumble motions\label{sec:five}}

A problem with the modelling of the rotational component of the bacterial run motion is the lack of an experimental turn-angle PDF. Nevertheless the experimental ratio $r_\mathrm{exp}$ between mean runtime and mean tumble time and the experimentally measured deflection angle \mbox{$\langle\abs{\psi}\rangle\pm \sigma_{\abs{\psi}}$} are available.
The fact that we can describe the Langevin dynamics of the turn deflection for both tumble and run using the control parameter $\beta$ makes us optimistic that it is possible to reconstruct the deflection PDF for the run motion on the base of these data. As the first step to this end, we use the ratio $r_\mathrm{exp}$ to find the theoretical deflection PDF of the run motion. In a second step, with the PDF we calculate theoretically the deflection angle and the its uncertainty and thus we confirm validity of our stochastic model for the direction changes of swimming bacteria.

\begin{figure}[ht]
\begin{center}
\includegraphics[scale=0.54]{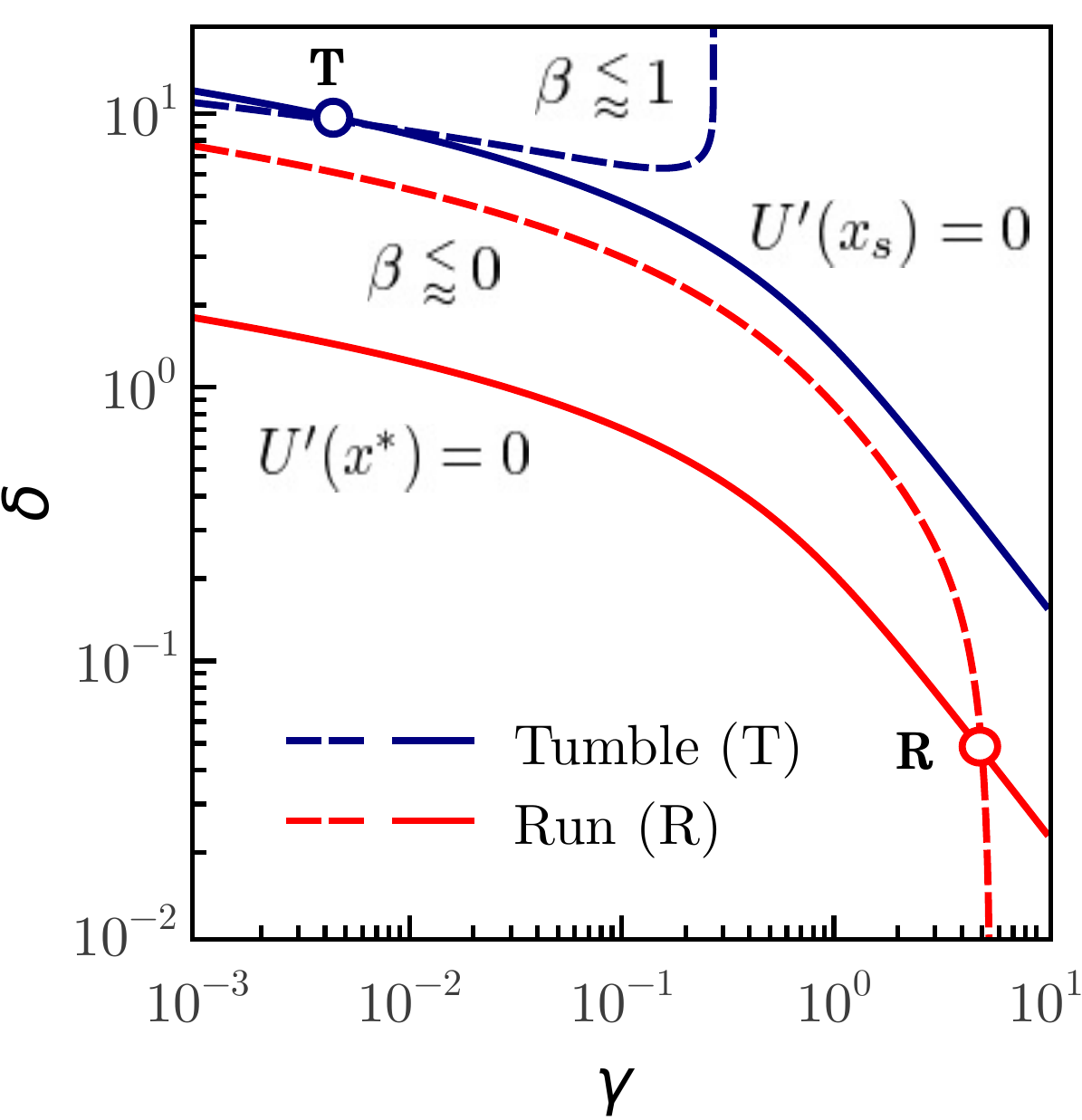} 
\end{center}
\caption{(Color online) In the parameter space $(\delta,\gamma)$, the point of intersection $\tumble$ (or $\run$) of the two-upper (or two-lower) curves determines the set of parameters that characterize the tumble (or run) motion. The dashed curves show the function $\delta=\delta(\beta,\gamma)$ the with control parameters $\beta\lessapprox 1$ and $\beta\lessapprox 0$ for the tumble and run motions, respectively. The upper solid curve shows the values for which $U'(x_\mathrm{s})=0$ (\textit{i.e.} $\gamma\sinh(\delta x_\mathrm{s})=1$), corresponding to the solution of stable equilibrium $x=x_\mathrm{s}$ of tumble motion. The lower solid curve shows the values for which $x=1$ is a global minimum of the potential $U$, which is equivalent to the use of a parameter $x^*>1$ such that $U'(x^*)=0$ (\textit{i.e.} $\gamma\sinh(\delta x^*)=1$), as can be seen in the plot shown on the left side of Figure~\ref{fig:run-motion-2}. In this plot we have used the following data: for the tumble $x_\mathrm{s}=0.63$ and $\beta=0.965$ and for the run $x^*=4.25$ and $\beta=-0.010$. With these data the obtained parameters are $\tumble\approxeq(6.63\times 10^{-3}, 9.062)$ and $\run\approxeq (4.98, 4.71\times 10^{-2})$. Additionally, using equation~(\ref{eq:rate}) we find $r\approxeq 6.15$, which is in very good agreement with the experimental value $r_\mathrm{exp}\approx 6.14$. 
\label{fig:delta-vs-gamma-RT}}
\end{figure}

The experimental ratio $r_\mathrm{exp}$ between the mean runtime $\langle t_\run\rangle$ and the mean tumble time $\langle t_\tumble\rangle$ ({\sl e.g.} $r_\mathrm{exp}\approx 6.14$ for \textit{E. coli} \cite{Berg1972}) allows to estimate a realistic value $\mu/D$, for a given set of parameters $\gamma$ and $\delta$ taking into account that $\mu$ is linked with $\delta$, $\gamma$ and the characteristic time $\tau$ {\sl via} equation~(\ref{tau}). Considering experimentally measured time distributions one can use the exponential distributions $P_\mathtt{J}(t)=\mathcal{N}_\mathtt{J}\,\exp[-\varepsilon\,t/(D_\mathrm{J}\tau_\mathrm{J})]$ (with $\mathtt{J}=\run, \tumble$ and $\varepsilon$ is a constant that makes exponent dimensionless) to obtain the ratio \mbox{$r=(D_\run\tau_\run)/(D_\tumble\tau_\tumble)$}, where the characteristic times are given by equation~(\ref{tau}). While for tumble motion the fit of our model to the experimental data yields that  \mbox{$\mu_\tumble = D_\tumble\,$}, that the run we propose \mbox{$\mu_\run =r\,D_\run$}. This leads to the relationship 
\begin{equation}
\delta_\tumble\,\sqrt{1+\gamma_\tumble^{\,2}}=r^2\,\delta_\run\,\sqrt{1+\gamma_\run^{\,2}}\label{eq:rate}
\end{equation}
with $\delta_\run=\delta(\beta_\run,\gamma_\run)$ (see equation~(\ref{eq:delta})), which is in agreement with the experimental data. The plot shown on the left side of Figure~\ref{fig:run-motion-2} shows the ratio $r$ as function of the parameter $\gamma_\run$. Assuming that $\gamma_\run\gg 1$ and $\gamma_\tumble\ll 1$, the theoretical $r$ has been calculated from \mbox{$\delta_\tumble\simeq r^2\,\delta_\run\,\gamma_\run$}. The validity of this approximation is supported by the value of the theoretical $r$ which is in good agreement with the experimental ratio $r_\mathrm{exp}\,$. Replacing $\mu=\rho\,D$ in equation~(\ref{ec1}) with $\delta=\delta(\beta,\gamma)$ (given by equation~(\ref{eq:delta})) the 3-parameter potential for the run and tumble motion is
\begin{equation}
U(x)=U_0-\rho\Bigl[x-\frac{\gamma}{\delta}\,\cosh(\delta x)\Bigr]\;,
\label{ec:pot2}
\end{equation}
with \mbox{$\rho=1$} for tumble motion and \mbox{$\rho=r$} for run motion. The plot shown on the right side of Figure~\ref{fig:run-motion-2} shows the potential $U(x)$ for the run motion on interval $\abs{x}\le 1$, where $x$ has a physical meaning (the turn angle is $\psi\in[0,2\pi)$) and outside of this interval where the parameter $x^*$ has no physical meaning (the angle $\psi$ is an imaginary number).  

\begin{figure}
\begin{center}
\includegraphics[scale=0.515]{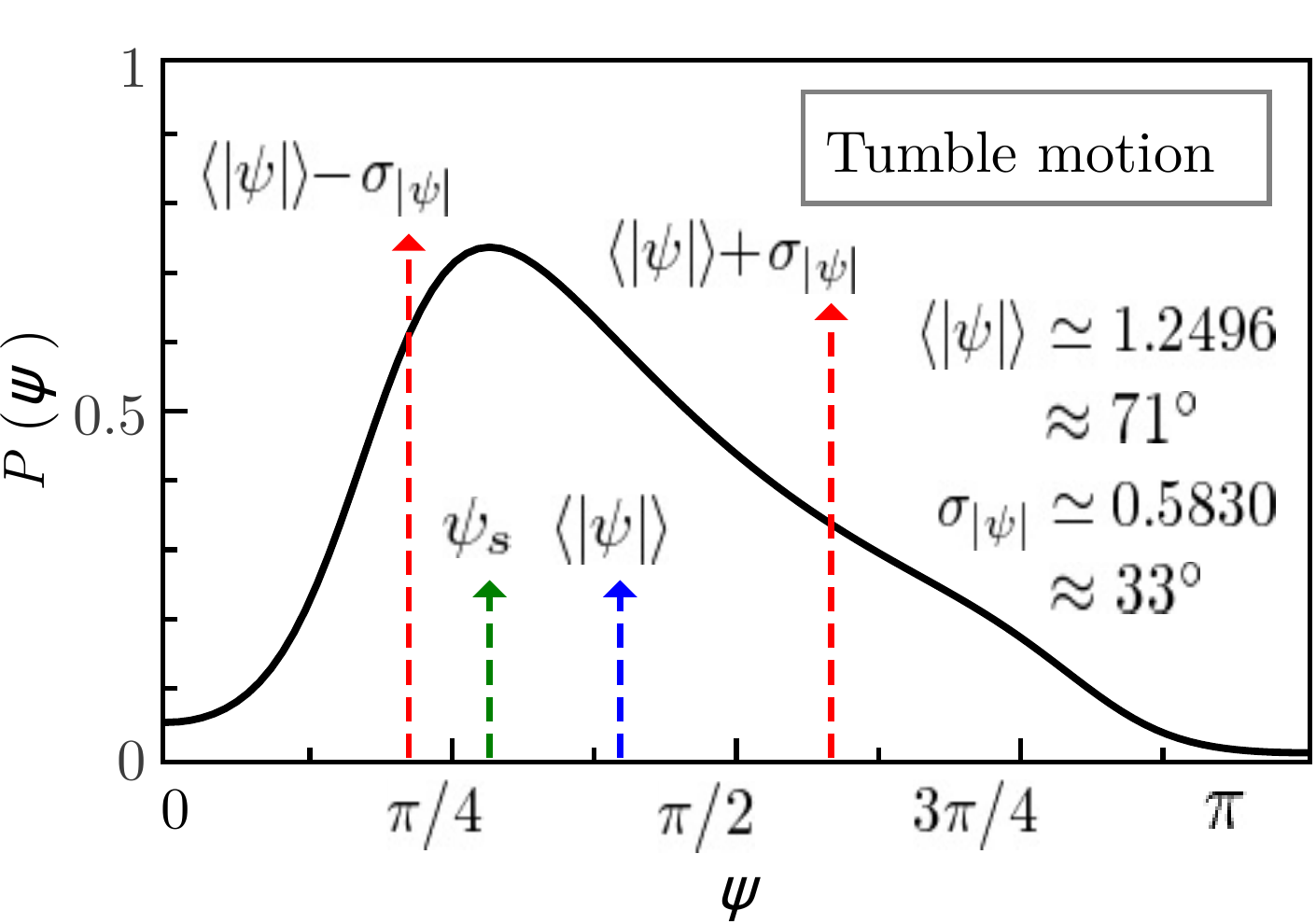}\hspace{1cm}
\includegraphics[scale=0.5]{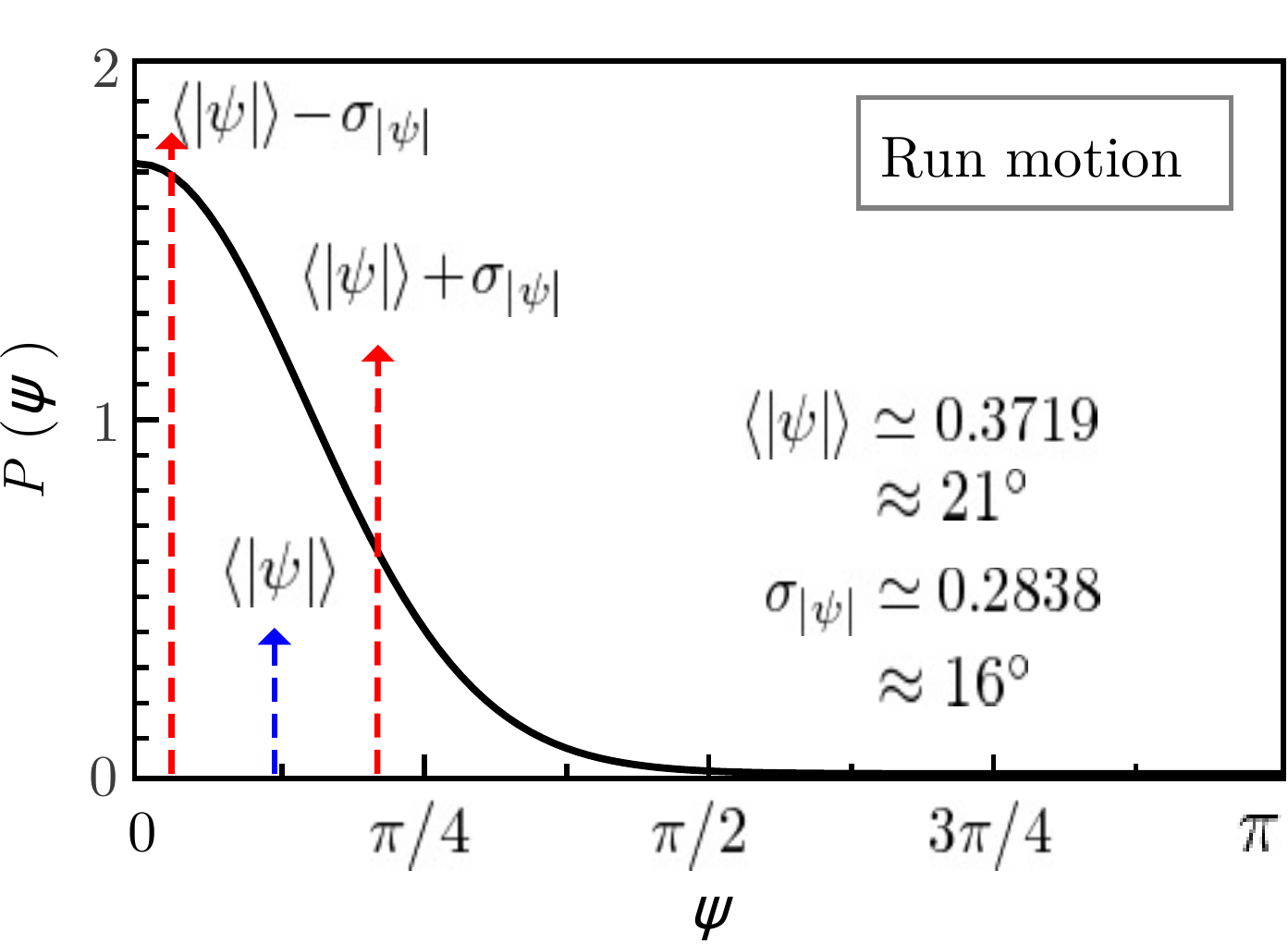}
\end{center}
\caption{(Color online) Both plots show the PDF of turn angles. Notice that the PDF is an even function of $\psi$. We used the same data as in Figure~\ref{fig:delta-vs-gamma-RT} with the purpose of showing for the two motions the high predictivity of the model. The mean and uncertainty of the angle absolute value $\abs{\psi}$ show in the plots are very close to the experimental data \cite{Berg1972}. In plot shown on the left side the most probable angle $\psi_\mathrm{s}\approxeq 51\degree$ corresponds to steady-state solution of the deflection $x_\mathrm{s}\approxeq 0.63\,$. \label{fig:run-and-tumble}}
\end{figure}

If the run-and-tumble dynamics are expresser by a single Langevin equation both  movements have to be linked. In the present work this linking is achieved by control parameter $\beta$ which clearly separates the solutions of both movements, with well-differentiated behaviours. Figure~\ref{fig:delta-vs-gamma-RT} visualizes the run-tumble transition mechanisms. At the points $\run$ (run) and $\tumble$ (tumble) of the parameter space $(\gamma, \delta)$ the system is in stable equilibrium and the control parameter $\beta$ takes a value $\beta_\run\lessapprox 0$ at point $\run$ and $\beta_\tumble\lessapprox 1$ at point $\tumble$. The critical control parameter $\beta_\mathrm{c}=0$ clearly separates run and tumble motions. Taking into account that the value of the control parameter for the tumble motion is $\beta_\tumble\lessapprox 1$, it is possible that the system approaches criticality by moving the control parameter $\beta\to 0^+$ along the equilibrium condition $U'(x_\mathrm{s})=0$ and that the noise eventually causes the loss of stability which leads to the transition $\tumble\to\run$\,. At the equilibrium point $\run$, the value of the control parameter $\beta_\run\lessapprox 0$ is already close to the critical value, whereby the noise destabilizes the equilibrium leading to the reverse transition $\run\to\tumble$\,.
These transition mechanisms can only be sustained if the effects of noise are more important for the tumble than for the run movement. Since the tumble time is smaller than the run time, the tumble motion can lose its stability more easily than the run motion. The proposed mechanisms should be complemented with stability studies of solutions in the presence of noise, which go beyond the scope of this paper. Finally, by moving the parameters of the system continuously we can transit from the turn-angles PDF of the run to the tumble and {\sl viceversa}. Figure~\ref{fig:run-and-tumble} shows both PDFs for values of parameters that reproduce the experimental observables of the run and tumble motions.

\section{Conclusiones}
More than four decades ago Berg and Brown \cite{Berg1972} measured the tumble-angle distribution (TAD) of swimming wild-type {\sl E. coli} bacteria which moved without taxis in an isotropic solution. On the base of this work, we build a stochastic model using a Langevin equation for the turn angles. We use the fact, that the normalized TAD can be represented as a series in $x=\cos\psi$ and study the deflection $x(t)$ as a stochastic process. We proposed drift and diffusion coefficients for the Fokker-Planck equation and the corresponding probability density function (PDF) reproduces the experimental TAD with very good agreement. Keeping the diffusion coefficient constant leads us to a stochastic process with additive noise and to a simple potential function $U(x)$ which is characterized by three fit parameters $\{\gamma,\delta,\mu\}$, where $\mu$ is directly linked to the noise intensity $D$. We give a physical meaning to these parameters showing that they are related to the characteristic time $\tau$, the steady state solution $x_\mathrm{s}$, and the control parameter $\beta$ of the tumble motion. We determine the Green's function associated with the Langevin equation of stochastic process $x$, taking advantage of the fact that the system is fully integrable in absence of noise. We show that the homogeneous contribution of the stochastic solution is related to the drift and the particular contribution is related to the noise. From this we conclude that the tumble motion is primarily caused by the flagellar motor and complementarily by diffusion. The contribution of the angular boost generated by the flagellar motor, which is related to the gradient of the potential $U(x)$, drives the system within stable equilibrium. On the other hand, the noise contribution to angular momentum moves the system away from the stable equilibrium state. The rotational diffusion, which is caused by the noise of the system, overlaps with the motion caused by the flagellar motor and determines width of tumble-angle PDF. 
We conclude from the covariance of the process that at very-long times scales the turn process becomes stationary, but at short time scales, which are on the order of the characteristic time of the tumble motion, the system is far from being stationary and the variance is time dependent. Assuming an exponential PDF of tumble time (referred to as TTD) we calculate the average variance of the tumble process. We use our model to show that small turns of bacteria around their centres of mass which occur during the run can be well modelled. Our model reveals that the stochastic turns during the run motion can be explained by an Ornstein-Uhlenbeck process, which for typical run times is stationary. This result confirms that during the run the rotational drift motions do not exist or are negligible and that only the rotational diffusion remains. 
In general we show that different turn movements of swimming \textit{E. coli} are characterized by a control parameter $\beta$ taking values $\beta_\tumble\lessapprox 1$ (for tumble motion) or $\beta_\run\lessapprox 0$ (for run motion). The control parameter $\beta$ determines the way that the system can make run-tumble transitions passing the critical value $\beta_\mathrm{c}=0$, while the system is in a stable equilibrium above and below to the critical value during the movement of tumble and run, respectively. Close to the criticality, the noise drives the system from a stable equilibrium (of run or tumble) to an unstable equilibrium from which it transits to a new stable equilibrium state (of tumble or run, respectively). On the basis of limited available experimental data we suggest a possible self-consistent model with high predictability, in which the parameters have a clear physical meaning. This work leaves several open questions, which have to be addressed in future studies. Taking into account that our model is only based on {\sl E. coli} data, it would be desirable to check whether this model is capable to describe the turning behaviour of other bacterial species in isotropic media without taxis. The consideration of external potentials in order to study systems with taxis is also pending. In order to investigate velocity correlations and the mean square displacement of the run and tumble motion  it is necessary to introduce Langevin equations for the speed and deflection of the bacterium and additionally take into account the present results. Finally, our conclusions about the run-tumble transition require new studies at a time scale which is much shorter than the characteristic tumble time during which biochemical processes in the flagellar motor occur. 

\section*{Acknowledgments}
This work was partially supported by Consejo Nacional de Investigaciones Cient\'{\i}ficas y T\'ecnicas (CONICET), Argentina, PIP 2014/16 N$^\circ$ 112-201301-00629.

\bibliography{BFH.swimming-biblio}

\end{document}